# A New Extended Mixture Model of Residual Lifetime Distributions

M. Kayid[1]

Dept.of Statistics and Operations Research,
College of Science, King Saud University,
P.O. Box 2455, Riyadh 11451, KSA

S. Izadkhah[2]

School of Mathematical Sciences
Ferdowsi University of Mashhad, Mashhad, Iran

[1]Email: drkayid@ksu.edu.sa
[2]Email: sa_ei924@stu-mail.um.ac.ir

**Abstract**

In this paper, we first propose a new extended mixture model of residual lifetime distributions. We show that this model is suitable in modeling residual lifetime in some practical situations. Several closure properties of some well-known dependence concepts, stochastic orders and aging notions under the formation of this model, are obtained. Finally, preservation properties of some stochastic orders under the formation of the model are discussed and some examples of interest are presented.

**Mathematics Subject Classification 2010**: 62N05, 62E10, 60K20

**Keywords**: Residual lifetime, Preservation, Mixture model, Likelihood ratio order, Hazard rate order, Mean residual life order, Characterization, Dependence, DLR, DFR, IMRL.

# 1 Introduction

Mixture models are widely used as computationally convenient representations for modeling complex probability distributions. In practical situations, it often happens that data from several populations are mixed and information about which subpopulation gave rise to individual data points is unavailable. Mixture models are used to model such data sets in nature. For example, measurements of life lengths of a device may be gathered without regard to the manufacturer, or data may be gathered on humans without regard, say, to blood type. If the ignored variable (manufacturer or blood type) has a bearing on the characteristic being measured, then the data are said to come from a mixture. Actually, it is hard to find data that are not some kind of a mixture, because there is almost always some relevant covariate that is not observed (cf. Barlow and Proschan [1], Marshall and Olkin [2], and the references therein). Let $\mathcal{F} = \{F(\cdot \mid \theta) : \theta \in \chi\}$ be a family of distributions indexed by a parameter $\theta$ which takes values in a set $\chi$. When $\theta$ can be regarded as a random variable with a distribution function $H$, then

$$F^*(x) = \int_\chi F(x \mid \theta) \, dH(\theta),$$

is the mixture of $\mathcal{F}$ with respect to $H$, and $H$ is called the mixing distribution. The corresponding survival function $\bar{F}^* = 1 - F^*$ is given by

$$\bar{F}^*(x) = \int_\chi \bar{F}(x \mid \theta) \, dH(\theta). \tag{1}$$

This model has been frequently used in the literature when $\bar{F}(x \mid \theta)$ stands for some semi parametric family of distributions where $\theta$ could not be considered as a constant (cf. Nanda and Das [3] and Gupta and Kirmani [4]).

Suppose now that the survival probability function of a fresh unit corresponding to a mission of duration $x$ is $\bar{F}(x) = 1 - F(x)$, where $F$ is the life distribution of the unit. According to Barlow and Proschan [1], the corresponding conditional survival of a unit of age $\theta$, when $\bar{F}(\theta) > 0$, is

$$\bar{F}(x \mid \theta) = \frac{\bar{F}(x + \theta)}{\bar{F}(\theta)}, \text{ for all } x \geq 0. \tag{2}$$

The model (2) is well-known in the literature as the family of residual lifetime distributions and the parameter $\theta$ is called age parameter (cf. Marshall and Olkin [2]). It is known that in many practical circumstances the parameter $\theta$ may not be constant due to various reasons, and the occurrence of heterogeneity is sometimes unpredictable and unexplained. The heterogeneity sometimes may not be possible to be neglected. To be specific, consider a population composed of lifetime devices of various ages that are still working. Suppose that a device is randomly taken from the population which its age is naturally unknown. For evaluating the residual life of this device after the time up which it has already survived, the parametric residual lifetime distribution with a constant parameter does not work. This is because the age of the selected device is indeed a random variable. Thus, in the model (2), it is important to investigate the influence of the random ages on the residual lifetime distribution.

The purpose of this paper is to propose a new extended mixture model of the family of residual lifetime distributions. In view of this model, several characterizations and closure properties of some dependence structures, stochastic orders and aging notions are established. We show that some stochastic orders between two random age variables are translated to the same stochastic orders between the average residual life variables. In addition, we demonstrate how the variation of the baseline variable with respect to some stochastic orders has an effect on the new mixture model. The rest of the paper is organized as follows. In Section 2, for ease of reference, we present some definitions and basic properties which will be used in the sequel. In Section 3, the new model and its representation are described. In that section, we provide some examples to describe the usefulness of the proposed model in practical situations. In addition, based on some reliability measures, we give some alternative representations for the new model. Closure properties of the model with respect to some dependence structures and some stochastic orders are studied in Section 4. In Section 5, we provide some preservation properties of a number of aging classes of life distributions under the formation of the model. In Section 6, in view of the proposed model, we establish some useful stochastic order relations. Finally in Section 7, we conclude the paper with some remarks of current research.

Throughout the paper, we use increasing and decreasing in place of non-decreasing and non-increasing, respectively. In addition, all the integrals and the expectations are assumed to exist when they are appeared.

## 2 Preliminaries

In reliability and survival studies, the hazard rate (HR), the reversed hazard rate (RHR) and the mean residual life (MRL) functions are very important measures. For the random variable $X$, the HR function is given by $r_X(x) = f(x)/\bar{F}(x)$, $x \geq 0$, the RHR function is given by $\widetilde{r}_X(x) = f(x)/F(x)$, $x > 0$ and the MRL function is given by $m_X(x) = \int_x^\infty \bar{F}(t)dt \ /\bar{F}(x)$, $x \geq 0$. In the following, we present definitions of some stochastic orders and aging notions used throughout the paper. For stochastic orders we refer to Shaked and Shanthikumar [5] and Nanda et al. [6] and for the aging notions we refer to Barlow and Proschan [1], Marshall and Olkin [2], Lai and Xie [7] and Righter et al. [8].

**Definition 2.1.**

Let $X$ and $Y$ be two nonnegative random variables with df's $F$ and $G$, sf's $\bar{F}$ and $\bar{G}$, pdf's $f$ and $g$, MRL functions $m_X$ and $m_Y$, HR functions $r_X$ and $r_Y$, and RHR functions $\widetilde{r}_X$ and $\widetilde{r}_Y$, respectively. We say that $X$ is smaller than $Y$ in the:

(i) Likelihood ratio order (denoted as $X \leq_{LR} Y$), if $g(x)/f(x)$, is increasing in $x > 0$.

(ii) Hazard rate order (denoted as $X \leq_{HR} Y$), if $r_X(x) \geq r_Y(x)$, for all $x \geq 0$.

(iii) Reversed hazard rate order (denoted as $X \leq_{RH} Y$), if $\widetilde{r}_X(x) \leq \widetilde{r}_Y(x)$, for all $x > 0$.

(iv) Aging intensity (denoted as $X \leq_{AI} Y$) if $\int_0^x r_X(u)du/ \int_0^x r_Y(u)du$ is increasing in $x > 0$.

(v) Usual stochastic order (denoted as $X \leq_{ST} Y$), if $\overline{F}(x) \leq \overline{G}(x)$, for all $x \geq 0$.

(vi) Mean residual life (denoted as $X \leq_{MRL} Y$), if $m_X(x) \leq m_Y(x)$, for all $x \geq 0$.

**Definition 2.2.**

Let $X$ and $Y$ be two nonnegative random variables. It said that $X$ is smaller than $Y$ in the upshifted likelihood ratio order (upshifted hazard rate order, up shifted mean residual life order), denoted as $X \leq_{LR\uparrow} (\leq_{HR\uparrow}, \leq_{MRL\uparrow})Y$, if

$$X - x \leq_{LR\uparrow} (\leq_{HR\uparrow}, \ \leq_{MRL\uparrow})Y \ \ \text{for all } x \geq 0.$$

For properties and applications of the upshifted stochastic orders we refer to Shaked and Shanthikumar [5].

**Definition 2.3.**

The nonnegative random variable $X$ is said to have:

(i) Increasing (decreasing) likelihood ratio property [ILR (DLR)], if $f$ is a log-concave (log-convex) function on $(0, \infty)$.

(ii) Increasing (decreasing) hazard rate property [IFR (DFR)], if $r_X$ is a increasing (decreasing) function on $(0, \infty)$.

(iii) Decreasing (increasing) mean residual life property [DMRL (IMRL)], if $m_X$ is a decreasing (increasing) function on $(0, \infty)$.

(iv) New better than used (new worse than used) [NBU (NWU)], if $\bar{F}(x+y) \leq (\geq) \bar{F}(x)\bar{F}(y)$, for all $x, y \geq 0$.

(v) New better than used in expectation (new worse than used in expectation) [NBUE (NWUE)], if $m_X(x) \leq (\geq) E(X)$, for all $x > 0$, provided that $X$ has a finite mean.

**Definition 2.4.** (Karlin [9])

A nonnegative function $\beta(x, y)$ is said to be totally positive (reverse regular) of order 2, denoted as $TP_2$ ($RR_2$), in $(x, y) \in \chi \times \gamma$, if

$$\begin{vmatrix} \beta(x_1, y_1) & \beta(x_1, y_2) \\ \beta(x_2, y_1) & \beta(x_2, y_2) \end{vmatrix} \geq (\leq) 0,$$

for all $x_1 \leq x_2 \in \chi$, and $y_1 \leq y_2 \in \gamma$, in which $\chi$ and $\gamma$ are two real subsets of the real line $\mathbb{R}$.

## 3 The mixture model

This section provides the main definition of the new model. In addition, several useful representations of the new model via some reliability measures are presented. Let $\Theta$ be a nonnegative random variable with df $H$, sf $\bar{H} = 1 - H$ and pdf $h$ whenever it exists. As mentioned before the age parameter $\theta$ in the family of residual lifetime distributions may not be constant. Thus, using a mixture distribution we extend the model (2) to a more general case. Formally, to handle the heterogeneity of the age parameter $\theta$ in residual lifetime family of distributions, we introduce the survival function of weighted average of residual lifetimes with respect to the mixing distribution $H$ as

$$\begin{aligned} \bar{F}^*(x) &= \int_0^\infty \left[ \bar{F}(x+\theta) / \bar{F}(\theta) \right] dH(\theta) \\ &= E\left[ \bar{F}(x+\Theta) / \bar{F}(\Theta) \right], \quad \text{for all } x \geq 0, \end{aligned} \quad (3)$$

where the expectation is taking with respect to $\Theta$. In the sequel, the random variable which has the sf $\bar{F}^*$ is denoted by $X^*$ with pdf $f^*$. Moreover, the random variables $X$, $\Theta$, and $X^*$ whose distributions were involved in (3), are called the baseline, the random age, and the average residual life variables, respectively. When the baseline distribution $F$ is absolutely continuous, the model can be stated as

$$\begin{aligned} f^*(x) &= \int_0^\infty \left[ f(x+\theta) \,/\, \bar{F}(\theta) \right] dH(\theta) \\ &= E\left[ f(x+\Theta) \,/\, \bar{F}(\Theta) \right], \text{ for all } x \geq 0. \end{aligned} \quad (4)$$

To illustrate the usefulness of the new model in practical situations we present the following examples.

**Example 3.1.**

Suppose that $X_1, X_2, ..., X_n$ denote the components lifetime of the system where $X_i$'s are assumed to be i.i.d. with a common absolutely continuous distribution function $F$ and density function $f$. If $X_{1:n}, X_{2:n}, ..., X_{n:n}$ represent the ordered lifetimes of the components, then in a system with signature vector $S = (s_1, ..., s_i, 0, ..., 0)$, the components with lifetime $X_{i+1:n}, ..., X_{n:n}$ would never cause the failure of the system. Hence, after the failure of the system, these components remain unfailed. Denote by $Y_j^{(i)}$, $j = 1, ..., n-i$, the randomly ordered values of $X_{j:n}$, $j = i+1, ..., n$. Then, the residual lifetime of the live components after the failure of the system can be denoted by

$$X_j^* = Y_j^{(i)} - \Theta, \quad j = 1, ..., n-i,$$

where $\Theta$ represents the lifetime of the system. According to Kelkin Nama et al. [10], the common marginal survival function of $X_j^*$ admits the form of the mixture model given in (3).

**Example 3.2.**

Suppose that $X_1, ..., X_n$ represent the lifetimes of the components of a $(n-k+1)$-out-of-$n$ system which are i.i.d. with common continuous df $F$ and pdf $f$. Denote by $X_1^{(k)}, ..., X_{n-k}^{(k)}$ the residual lifetimes of the components after the $k$ failures in the system. Bairamov and Arnold [11] have studied some distributional properties of $X_j^{(k)}$'s, $j = 1, ..., n-k$. The joint sf of $X_j^{(k)}$'s is given by

$$\bar{F}_n^{(k)}(x_1, ..., x_{n-k}) = \int_0^\infty \left\{ \prod_{j=1}^{n-k} \left[ \bar{F}(t+x_j) \,/\, \bar{F}(t) \right] \right\} dF_{k:n}(t),$$

where $F_{k:n}$ denotes the distribution of the $k$th order statistic $X_{k:n}$. The marginal sf of $X_j^{(k)}$, $j = 1, ..., n-k$ can be derived as

$$\bar{F}^*(x_j) = \int_0^\infty \left[ \bar{F}(x_j + t) \,/\, \bar{F}(t) \right] dF_{k:n}(t)$$

which coincides with the new model given in (3).

**Remark 3.1.**

It is important to notice that in the cases that studied in the Example 3.2, the baseline variable is the lifetime of components, the random age variable $\Theta$ is the $k$th order statistic and the average residual life variable $X^*$ is the residual lifetime of the alive components after the failure of the system. As a special case, the new model with the baseline variable $X_1$ and the random age variable $X_{n-1:n}$ gives the average residual life $X_{n:n} - X_{n-1:n}$ which is the last sample spacing arising from the random sample $X_1, ..., X_n$.

**Example 3.3.**

Consider a population including some used devices (that are still at work) with different ages $\theta_1, \theta_2$, etc., say. Therefore, we have a mixture formed population in which various used devices with different ages are mixed. The age parameter $\theta$ is not constant in this population because it varies from one used device to another one. Thus, in this population we have a random age parameter $\Theta$. As a result, the random variable $X^*$ can be used to model the average residual lifetime of the devices in the total population.

In the rest of this section, based on some reliability measures, we give some alternative representations for the new model. Denote by $\Pi(\cdot \mid x)$ the df of the random variable $(\Theta \mid X^* > x)$ which, for all $\theta > 0$ and for any $x \geq 0$, is given by

$$\Pi(\theta \mid x) = \int_0^\theta \left[ \bar{F}(x+w) \ / \ \bar{F}(w)\bar{F}^*(x) \right] \, dH(w). \tag{5}$$

To see how the HR of $X$ and the HR of $X^*$ are connected to each other we get

$$r_{X^*}(x) = \int_0^\infty \left[ f(x+\theta) \ / \ \bar{F}(\theta)\bar{F}^*(x) \right] dH(\theta)$$

$$= \int_0^\infty \left[ f(x+\theta) \ / \ \bar{F}(x+\theta) \right] d\Pi(\theta \mid x)$$

$$= E\left[ r_X(x+\Theta) \mid X^* > x \right], \text{ for all } x \geq 0. \tag{6}$$

In addition, the MRL functions of $X^*$ and $X$ are connected as

$$m_{X^*}(x) = \int_x^\infty \int_0^\infty \left[ \bar{F}(t+\theta) \ / \ \bar{F}(\theta)\bar{F}^*(x) \right] dH(\theta) dt$$

$$= \int_0^\infty \int_{x+\theta}^\infty \left[ \bar{F}(t) \ / \ \bar{F}(x+\theta) \right] dt \, d\Pi(\theta \mid x)$$

$$= E\left[ m_X(x+\Theta) \mid X^* > x \right], \text{ for all } x \geq 0. \tag{7}$$

## 4 Dependence, characterization and closure properties

In this section, we first show that the random variables $X^*$ and $\Theta$ satisfy some dependence structures depending on aging properties of the baseline variable $X$. Then, some characterizations of aging properties are developed in view of the new model. Under some assumptions, we establish that the model enjoys from some closure properties with respect to several stochastic orders. In what follows we define some well-known dependence concepts according to Nelsen [12].

**Definition 4.1.**

Let $X^*$ and $\Theta$ have the common support $(0, \infty)$ and let $(X^*, \Theta)$ have the joint sf $\bar{H}$ and the joint pdf $h$. The random variables $X^*$ and $\Theta$ are said to have:

(i) Positive (negative) likelihood ratio dependence structure [PLRD (NLRD)] if $h(x,\theta)$ is $TP_2$ ($RR_2$) in $(x,\theta) \in (0,\infty) \times (0,\infty)$.

(ii) Stochastically increasing (decreasing) property of $X^*$ in $\Theta$ [SI($X^* \mid \Theta$) (SD($X^* \mid \Theta$))] if $P(X^* > x \mid \Theta = \theta)$ is increasing (decreasing) in $\theta$, for all $x \in (0, \infty)$.

(iii) Right corner set increasing (decreasing) property [RCSI (RCSD)] if $\bar{H}(x, \theta)$ is TP$_2$ ( RR$_2$) in $(x, \theta) \in (0, \infty) \times (0, \infty)$.

Now, we have the following result.

**Theorem 4.1.**

Let $X^*$ and $\Theta$ be as described in Definition 4.1. Then

(i) $X^*$ and $\Theta$ are PLRD (NLRD) if, and only if $X$ is DLR (ILR).

(ii) SI($X^* \mid \Theta$) (SD($X^* \mid \Theta$)) if, and only if $X$ is DFR (IFR).

(iii) $X^*$ and $\Theta$ are RCSI (RCSD) if $X$ is DFR (IFR).

**Proof.**

(i). The joint pdf of $(X^*, \Theta)$, for all $x, \theta \geq 0$ is given by

$$h(x, \theta) = f(x \mid \theta) h(\theta)$$

$$= \left[ f(x + \theta) / \bar{F}(\theta) \right] h(\theta),$$

where $f(x \mid \theta)$ is the conditional density of $X^*$ given that $\Theta = \theta$. It is well-known that $X$ is DLR (ILR) if, and only if $f(x + \theta)$ and also $h(x, \theta)$ is TP$_2$ (RR$_2$) in $(x, \theta) \in (0, \infty) \times (0, \infty)$. Hence the proof of (i) is completed.

(ii). Observe that, for all $x, \theta \geq 0$

$$P(X^* > x \mid \Theta = \theta) = \bar{F}(x + \theta) / \bar{F}(\theta),$$

which is increasing (decreasing) in $\theta$, for all $x \geq 0$, if and only if $X$ is DFR (IFR).

(iii). For all $x, \theta \geq 0$, we have

$$\bar{H}(x, \theta) = \int_0^\infty \left\{ \left[ \bar{F}(x + w) / \bar{F}(w) \right] \times I(w - \theta) \right\} dH(w),$$

where $I(x) = 0$ when $x < 0$, and $I(x) = 1$ when $x \geq 0$. Since $X$ is DFR (IFR), $\bar{F}(x+w)/\bar{F}(w)$ is TP$_2$ (RR$_2$) in $(x, w)$. In addition, it is easy to see that $I(w - \theta)$ is TP$_2$ in $(w, \theta)$. Applying the general composition theorem (Lemma 1.1, p. 99) of Karlin [9] to the above identity, we conclude that $\bar{H}(x, \theta)$ is TP$_2$ (RR$_2$) in $(x, \theta)$, which completes the proof. ∎

**Remark 4.1.**

As a useful conclusion of Theorem 4.1, if $X$ has "no-aging" property, i.e. if $X$ has the exponential distribution, then $X^*$ and $\Theta$ in the new model are independent and vise versa.

In view of Theorem 4.1 and Example 3.2 the following corollary is immediate.

**Corollary 4.1.**

The following assertions hold:

(i) $X_{n:n} - X_{n-1:n}$ and $X_{n-1:n}$ are PLRD (NLRD) if and only if $X_1$ is DLR (ILR).

(ii) SI($X_{n:n} - X_{n-1:n} \mid X_{n-1:n}$) (SD($X_{n:n} - X_{n-1:n} \mid X_{n-1:n}$)) if and only if $X_1$ is DFR (IFR).

(iii) If $X_1$ is DFR (IFR), then $X_{n:n} - X_{n-1:n}$ and $X_{n-1:n}$ are RCSI (RCSD).

As a particular case, in view of Theorem 4.1(i), we have:

**Corollary 4.2.**

Let $X_1, ..., X_n$ be a random sample of continuous random variables with support $(0, \infty)$ and let $X_{k:n}$ denote the $k$th order statistic. Then, $X_{n:n} - X_{n-1:n}$ and $X_{n-1:n}$ are independent if, and only if $X_1$ is exponentially distributed.

In the following result, we provide some interesting characterizations.

**Theorem 4.2.**

Let $X$ be a lifetime random variable. Then:

(i) $X$ is NBU (NWU), if and only if, $X^* \leq_{ST} (\geq_{ST}) X$, for all nonnegative variables $\Theta$.

(ii) $X$ is IFR (DFR), if and only if, $X^* \leq_{HR} (\geq_{HR}) X$, for all nonnegative variables $\Theta$.

(iii) $X$ is ILR (DLR), if and only if, $X^* \leq_{LR} (\geq_{LR}) X$, for all nonnegative variables $\Theta$.

(iv) $X$ is NBUE (NWUE), if and only if, $E(X^*) \leq (\geq) E(X)$, for all nonnegative variables $\Theta$.

**Proof.**

(i) We know that $X$ is NBU (NWU) if, and only if $(X - \theta \mid X > \theta) \leq_{ST} (\geq_{ST}) X$, for all $\theta > 0$ (cf. Shaked and Shanthikumar [5]). This implies that, for all $t \geq 0$

$$\bar{F}(t) - \bar{F}^*(t) = E\left[\bar{F}(t) - \bar{F}(t + \Theta) / \bar{F}(\Theta)\right] \geq (\leq) 0.$$

Conversely, let $X^* \leq_{ST} (\geq_{ST}) X$. Then by taking $\Theta$ such that $P(\Theta = \theta) = 1$, we get $X$ is NBU (NWU).

(ii) Let $X$ be IFR (DFR). Then, in view of (6), for all $t \geq 0$

$$r_{X^*}(t) - r_X(t) = E\left[r_X(t + \Theta) - r_X(t) \mid X^* > x\right]$$

$$\geq (\leq) 0.$$

In the reversed direction, if we take $\Theta$ as a degenerate random variable we obtain $X^* \leq_{HR} (\geq_{HR}) X$ which implies that $(X - \theta \mid X > \theta) \leq_{HR} (\geq_{HR}) X$, for all $\theta > 0$. This is equivalent to saying that $X$ is IFR (DFR).

(iii) It is known that if $X$ is ILR (DLR), then $f(t + \theta)/f(t)$ is decreasing (increasing) in $t$, for all $\theta > 0$. So

$$f^*(t) / f(t) = E\left[f(t + \Theta) / f(t)\bar{F}(\Theta)\right],$$

is decreasing (increasing) in $t$. Conversely, if we take $\Theta$ such that $P(\Theta = \theta) = 1$, for each $\theta > 0$, then $X^* \leq_{LR} (\geq_{LR}) X$ gives $(X - \theta \mid X > \theta) \leq_{LR} (\geq_{LR}) X$, for all $\theta > 0$. That is $X$ is ILR (DLR).

(iv) Put $x = 0$ in (7), we get $E(X^*) = E[m_X(\Theta)]$. It is known that $X$ is NBUE (NWUE) if, and only if $E(X) \geq (\leq) m_X(\theta)$, for all $\theta > 0$. Hence, it follows that

$$E(X) - E(X^*) = E[E(X) - m_X(\Theta)]$$

$$\geq (\leq) 0.$$

To prove the converse, take $\Theta$ as a degenerate variable and as before the result is concluded. ∎

In the following theorem, we show that the new model is closed under some up shifted stochastic orders when appropriate assumptions are satisfied.

**Theorem 4.3.**

Let $X$ be a lifetime random variable. Then:

(i) If $X$ has ILR property, then $X^* \leq_{LR\uparrow} X$.

(ii) If $X$ has IFR property, then $X^* \leq_{HR\uparrow} X$.

(iii) If $X$ has DMRL property, then $X^* \leq_{MRL\uparrow} X$.

**Proof.**

(i) First, observe that if $X$ is ILR, then $f(t+x+\theta)/f(t)$ is decreasing in $t$, for all $x, \theta \geq 0$. Thus the ratio
$$f^*(t+x) \,/\, f(t) = E\left[f(t+x+\Theta) \,/\, f(t)\bar{F}(\Theta)\right],$$
is decreasing in $t$, for all $x \geq 0$. That is $X^* \leq_{LR\uparrow} X$.

(ii) Note that if $X$ is IFR, then $\bar{F}(t+x+\theta)/\bar{F}(t)$ is decreasing in $t$, for all $x, \theta \geq 0$. It follows that
$$\bar{F}^*(t+x) \,/\, \bar{F}(t) = E\left[\bar{F}(t+x+\Theta) \,/\, \bar{F}(t)\bar{F}(\Theta)\right],$$
is decreasing in $t$, for all $x \geq 0$, which is equivalent to $X^* \leq_{HR\uparrow} X$.

(iii) In view of Fubini theorem we get, for all $t, x \geq 0$,
$$\int_{t+x}^{\infty} \bar{F}^*(u)du = \int_{t+x}^{\infty} \int_{0}^{\infty} \bar{F}(u+\theta) \,/\, \bar{F}(\theta) \, dH(\theta)du$$
$$= E\left[\int_{t+x+\Theta}^{\infty} \bar{F}(u)du \,/\, \bar{F}(\Theta)\right].$$

The condition that $X$ is DMRL implies that $\int_{t+x+\theta}^{\infty} \bar{F}(u)du \,/\, \int_{t}^{\infty} \bar{F}(u)$ is decreasing in $t$, for all $x, \theta \geq 0$. Therefore,
$$\int_{t+x}^{\infty} \bar{F}^*(u)du \,/\, \int_{t}^{\infty} \bar{F}(u)du = E\left[\int_{t+x+\Theta}^{\infty} \bar{F}(u)du \,/\, \bar{F}(\Theta) \int_{t}^{\infty} \bar{F}(u)du\right],$$
is decreasing in $t$, for all $x \geq 0$, which completes the proof. ∎

**Remark 4.2.**

Let $X$ have a finite mean. Denote by $\widetilde{X}$ the random variable that has distribution $\widetilde{F}(x) = \int_{0}^{x} \bar{F}(u)du/E(X)$, $x \geq 0$, which is well-known in the literature as the equilibrium distribution associated with $F$. If the random age $\Theta$ in the new model is degenerate at $\theta$ and if $\Theta$ has df $\widetilde{F}$, each in one time, then the average residual life variable $X^*$ is equal in distribution with $X_\theta = (X - \theta \mid X > \theta)$ and $\widetilde{X}$, respectively.

The following conclusion is immediate from Theorem 4.3.

**Corollary 4.3.**

Let $X$ be ILR (IFR) [DMRL]. Then

(i) $X_\theta \leq_{LR\uparrow(HR\uparrow)[MRL\uparrow]} X$, for all $\theta > 0$.

(ii) $\tilde{X} \leq_{LR\uparrow(HR\uparrow)[MRL\uparrow]} X$.

(iii) $X_{n:n} - X_{n-1:n} \leq_{LR\uparrow(HR\uparrow)[MRL\uparrow]} X$.

In the following result, we establish closure property of the new model with respect to the aging intensity order when appropriate assumptions are imposed.

**Theorem 4.4.**

Let the HR function of the random variable $X$ be decreasing and log-concave. Then, $X \leq_{AI} X^*$.

**Proof.**

First, note that $X \leq_{AI} X^*$ if, and only if

$$\int_0^x [r_{X^*}(u)r_X(x) - r_{X^*}(x)r_X(u)]\, du \geq 0, \text{ for all } x > 0,$$

which simply holds if $r_{X^*}(x)/r_X(x)$ is decreasing in $x$. By (6) we have

$$r_{X^*}(x) / r_X(x) = E[r_X(x+\Theta) / r_X(x) \mid X^* > x]$$

$$= \int_0^\infty \phi(x,w)\, \pi(w \mid x)\, dw$$

$$= E[\phi(x,W)],$$

where $W$ is a nonnegative random variable with df given in (5) with the following pdf

$$\pi(w \mid x) = [\bar{F}(x+w) / \bar{F}(w)\bar{F}^*(x)]\, h(w), \quad w \geq 0,$$

for all $x \geq 0$, and $\phi(x,w) = r(x+w)/r(x)$ which is decreasing in $x$ and also it is decreasing in $w$. On the other hand, since $X$ is DFR, then $\bar{F}(x+w)$ is TP$_2$ in $(x,w)$ which implies that $W$ is increasing in $x$ with respect to the likelihood ratio order. That is $W$ is also stochastically increasing in $x$. Appealing to Lemma 2.2(i) of Misra and Van Der Meulen [12] the result is obtained. ∎

The following counterexample shows that the decreasing condition in Theorem 4.4 cannot be dropped.

**Counterexample 4.1.**

Let $X$ have Weibull distribution with survival function $\bar{F}(t) = e^{-t^2}, t \geq 0$. The HR function of $X$ is not decreasing but its log-concave. Let $\Theta$ be such that $P(\Theta = 0) = 0.25$ and $P(\Theta = 1) = 0.75$. Then, according to (3) the random variable $X^*$ has survival function $\bar{F}^*(t) = 0.25e^{-t^2} + 0.75e^{-t(t+2)}, t \geq 0$. It is easy to check that

$$\ln[\bar{F}(t)] / \ln[\bar{F}^*(t)] = t^2 [t^2 + \ln(4) - \ln(1 + 3e^{-2t})]^{-1},$$

is not an increasing function in $t \geq 0$. By Theorem 3.1(iii) of Nanda et al. [6] we deduce that $X \not\leq_{AI} X^*$.

The next counterexample reveals that the log-concavity condition in Theorem 4.4 cannot be dropped.

**Counterexample 4.2.**

Let $X$ have sf $\bar{F}(t) = 0.25e^{-t} + 0.75e^{-2t}$, $t \geq 0$. Then the HR function of $X$ is obtained as $r_X(t) = 2/(3+e^t)$, which is decreasing but not log-concave. If $\Theta$ is such that $P(\Theta = 1) = 1$, then after deriving $\bar{F}^*$ via (3) we have

$$\frac{\ln\left[\bar{F}(t)\right]}{\ln\left[\bar{F}^*(t)\right]} = \frac{\ln(e^{-t} + 3e^{-2t}) - \ln(4)}{\ln(e^{-(t+1)} + 3e^{-2(t+1)}) - \ln(e^{-1} + 3e^{-2})},$$

which is not an increasing function and hence $X \nleq_{AI} X^*$.

In view of Theorem 4.4., as particular cases, we derive the following corollary.

**Corollary 4.4.**

Suppose that $X$ is DFR such that its hazard rate is log-concave. Then

(i) $X \leq_{AI} X_\theta$, for all $\theta > 0$.

(ii) $X \leq_{AI} \widetilde{X}$.

(iii) $X \leq_{AI} X_{n:n} - X_{n-1:n}$.

# 5 Aging properties

In this section, we discuss preservation properties of some aging notions under the transformation $X \to X^*$ in the new model. The first result deals with the DLR aging property.

**Theorem 5.1.**

If $X$ is DLR, then $X^*$ is DLR.

**Proof.**

We need to show that $f^*(t+x)/f^*(x)$ is increasing in $x$, for all $t \geq 0$. First, note that

$$\begin{aligned} f^*(t+x)\,/\,f^*(x) &= E\left[f(t+x+\Theta)\,/\,\bar{F}(\Theta)\right]\,/\,E\left[f(x+\Theta)\,/\,\bar{F}(\Theta)\right] \\ &= \frac{\int_0^\infty \left\{[f(t+x+w)\,/\,f(x+w)] \times [f(x+w)\,/\,\bar{F}(w)]\right\} dH(w)}{\int_0^\infty \left[f(x+w)\,/\,\bar{F}(w)\right] dH(w)} \\ &= E(\phi(x,W)), \end{aligned}$$

where $\phi(x,w) = f(t+x+w)\,/\,f(x+w)$ and $W$ is a nonnegative random variable with the pdf

$$h(w \mid x) = \frac{[f(x+w)h(w)]\,/\,\bar{F}(w)}{\int_0^\infty \left[f(x+w)\,/\,\bar{F}(w)\right] dH(w)}, \quad w, x \geq 0.$$

Since $X$ is DLR, $\phi(x,w)$, for all $t \geq 0$, is increasing in either one of $x$ and $w$, when the other one is fixed. Moreover, the DLR property of $X$ implies that $\beta$ is log-convex i.e., $f(x+w)$ is TP$_2$ in $(x,w)$ which by known properties of TP$_2$ functions, it follows that $h(w \mid x)$ is also TP$_2$ in $(x,w)$. This is equivalent to the fact that $W$ is increasing in $x$ with respect to the likelihood ratio order and hence $W$ is stochastically increasing in $x$. Now, an application of Lemma 2.2(i) of Misra and Van Der Meulen [13] provides that $E\left[\phi(x,W)\right]$ is increasing in $x$, which concludes the proof. ∎

In the next theorem, the preservation property of the DFR class is obtained.

**Theorem 5.2.**

If $X$ is DFR, then $X^*$ is DFR.

**Proof.**

The proof will be validated if we prove that $\bar{F}^*(t+x)/\bar{F}^*(x)$ is increasing in $x$, for all $t \geq 0$. By (3), we have

$$\bar{F}^*(t+x) \,/\, \bar{F}^*(x) = E\left[\bar{F}(t+x+\Theta) \,/\, \bar{F}(\Theta)\right] \,/\, E\left[\bar{F}(x+\Theta) \,/\, \bar{F}(\Theta)\right]$$

$$= \frac{\int_0^\infty \left\{\left[\bar{F}(t+x+w) \,/\, \bar{F}(x+w)\right] \times \left[\bar{F}(x+w) \,/\, \bar{F}(w)\right]\right\} dH(w)}{\int_0^\infty \left[\bar{F}(x+w) \,/\, \bar{F}(w)\right] dH(w)}$$

$$= E\left[\phi(x, W)\right],$$

where $\phi(x, w) = \bar{F}(t+x+w)/\bar{F}(x+w)$ and $W$ is a nonnegative random variable with density

$$h(w \mid x) = \left[\int_0^\infty \left[\bar{F}(x+w) \,/\, \bar{F}(w)\right] dH(w)\right]^{-1} \times \left\{\left[\bar{F}(x+w)h(w)\right] \,/\, \bar{F}(w)\right\}, \ w, x \geq 0.$$

Since $X$ is DFR, then $\phi(x, w)$, for all $t \geq 0$, is increasing in $x$ and in $w$, when the other is fixed. The DFR property of $X$ means that $\bar{F}$ is log-convex i.e., both $\bar{F}(x+w)$ and $h(w \mid x)$ are $\text{TP}_2$ in $(x, w)$ and as in the proof of Theorem 5.1, $W$ is stochastically increasing in $x$. Applying Lemma 2.2(i) of Misra and Van Der Meulen [13] we deduce that $E\left[\phi(x, W)\right]$ is increasing in $x$, or equivalently $X^*$ is DFR. ∎

The following theorem states the preservation property of the IMRL class.

**Theorem 5.3.**

If $X$ is IMRL, then $X^*$ is IMRL.

**Proof.**

We need to show that $\int_{t+x}^\infty \bar{F}^*(u)du \,/\, \int_x^\infty \bar{F}^*(u)du$ is increasing in $x$, for all $t \geq 0$. Set $\nu(x) = \int_x^\infty \bar{F}(u)du$. As in the proof of Theorem 4.3(iii), for any $x, t \geq 0$ we can derive

$$\int_{t+x}^\infty \bar{F}^*(u)du \,/\, \int_x^\infty \bar{F}^*(u)du = E\left[\nu(t+x+\Theta) \,/\, \bar{F}(\Theta)\right] \,/\, E\left[\nu(x+\Theta) \,/\, \bar{F}(\Theta)\right]$$

$$= \frac{\int_0^\infty \left(\left[\nu(t+x+w) \,/\, \nu(x+w)\right] \times \left[\nu(x+w) \times \bar{F}(w)\right]\right) dH(w)}{\int_0^\infty \left[\nu(x+w) \,/\, \bar{F}(w)\right] dH(w)}$$

$$= E[\phi(x, W)],$$

where

$$\phi(x, w) = \nu(t+x+w) \,/\, \nu(x+w)$$

$$= \int_{t+x+w}^\infty \bar{F}(u)du \,/\, \int_{x+w}^\infty \bar{F}(u)du,$$

which by the assumption, for all $t \geq 0$, is increasing in $x$ and in $w$, whenever the other is fixed. The random variable $W$ has pdf

$$h(w \mid x) = \frac{[\nu(x+w)h(w)] \ / \ \bar{F}(w)}{\int_0^\infty [\nu(x+w) \ / \ \bar{F}(w)] \, dH(w)}, \text{ for all } w, x \geq 0.$$

Let $H(\cdot \mid x)$ be the df of $W$. Let us observe that

$$X \text{ is } IMRL \Leftrightarrow \nu(x+w) \text{ is } TP_2 \text{ in } (x,w)$$

$$\Leftrightarrow h(w \mid x) \text{ is } TP_2 \text{ in } (x,w)$$

$$\Leftrightarrow h(w \mid x_2) \ / \ h(w \mid x_1) \text{ is increasing in } w, \text{ for all } 0 \leq x_1 \leq x_2.$$

Because of the likelihood ratio order implies the usual stochastic order, by the above equivalence relations we deduce that $H(w \mid x_1) \geq H(w \mid x_2)$, for all $w \geq 0$, and for all $x_1 \leq x_2$. Again, appealing to Lemma 2.2(i) of Misra and Van Der Meulen [13] the result follows. ∎

**Corollary 5.1.**

Let $X$ be DLR (DFR) [IMRL]. Then $X_\theta$, for all $\theta > 0$, $\widetilde{X}$, and $X_{n:n} - X_{n-1:n}$ are DLR (DFR) [IMRL].

In the Theorems 5.1 - 5.3, we have established only the preservation properties of some negative aging classes. The following counterexample shows that the positive aging classes are not closed under the formation of the new model.

**Counterexample 5.1.**

Let $X$ have the sf $\bar{F}(t) = e^{-5t^2}$, $t \geq 0$, and let $\Theta$ have the sf $\bar{H}(\theta) = e^{-\theta}, \theta \geq 0$. Hence, the random variable $X$ has ILR, IFR and DMRL properties. Appealing to (3), the sf of $X^*$ can be obtained as

$$\bar{F}^*(t) = e^{-5t^2} \ / \ (1+10t), \ t \geq 0.$$

We observe that the MRL function of $X^*$ is not decreasing since $m_{X^*}(0.002) \cong 0151961$ and $m_{X^*}(0.004) \cong 0.15293$. Hence, $X^*$ does not have the DMRL property. Since the IFR and the ILR classes are subclasses of the DMRL class (cf. Lai and Xie [7]) thus the IFR and the ILR properties also do not satisfy for the random variable $X^*$.

# 6 Stochastic order relations

In this section, by some stochastic orders we study the influence of the variation of the random age variable and the variation of the baseline variable on the variation of the average residual life variable in the model. Let $\Theta_1$ and $\Theta_2$ be two nonnegative random variables and let the random variable $X_i^*$, for $i = 1, 2$, have the sf

$$\bar{F}_i^*(x) = E\left[\bar{F}(x+\Theta_i) \ / \ \bar{F}(\Theta_i)\right], \text{ for all } x \geq 0. \tag{8}$$

Hereafter, we assume that $\Theta_1$ and $\Theta_2$ have pdf's (df's) $h_1$ ($H_1$) and $h_2$ ($H_2$), respectively. Furthermore, we assume that $\Theta_1$ and $\Theta_2$ are independent.

**Theorem 6.1.**

Let $X$ be:

(i) DLR (ILR). Then $\Theta_1 \leq_{LR} \Theta_2$ implies $X_1^* \leq_{LR} (\geq_{LR}) X_2^*$.

(ii) DFR (IFR). Then $\Theta_1 \leq_{HR} \Theta_2$ implies $X_1^* \leq_{HR} (\geq_{HR}) X_2^*$.

**Proof.**

Under the condition given in (i) the conditional pdf of $(X^* \mid \Theta = \theta)$ is $TP_2$ ($RR_2$) in $(x, \theta)$ and under the stated condition in (ii) the conditional HR of $(X^* \mid \Theta = \theta)$ is decreasing (increasing) in $\theta$. Now, the proof will be obtained using Theorems 3.3 and 3.4 of Gupta and Gupta [14].

**Theorem 6.2.**

Let $X$ be DFR (IFR). Then

$$\Theta_1 \leq_{ST} \Theta_2 \Rightarrow X_1^* \leq_{ST} (\geq_{ST}) X_2^*.$$

In particular, if $X$ is IMRL (DMRL), then

$$\Theta_1 \leq_{ST} \Theta_2 \Rightarrow E(X_1^*) \leq (\geq) E(X_2^*).$$

**Proof.**

For the first part of the theorem, we prove the result when $X$ is DFR. The IFR case is similar. Note that if $X$ is DFR, then $\bar{F}(w+x)/\bar{F}(w)$ is increasing in $w$, for all $x \geq 0$. On the other hand, $\Theta_1 \leq_{ST} \Theta_2$ implies that

$$\int_\theta^\infty d\left[H_2(w) - H_1(w)\right] \geq 0, \text{ for all } \theta \geq 0. \tag{9}$$

By Lemma 7.1(a) of Barlow and Proschan [1] we can get

$$\int_0^\infty \left[\bar{F}(w+x) / \bar{F}(w)\right] d\left[H_2(w) - H_1(w)\right] \geq 0,$$

which means that $X_1^* \leq_{ST} X_2^*$. For the second part of the theorem, we give the proof for the case where $X$ is IMRL. The DMRL case is similar. From (7) we get $E(X_i^*) = E\left[m(\Theta_i)\right]$, for each $i = 1, 2$. The assumption that $X$ is IMRL implies that $m_X(w)$ is increasing in $w \geq 0$. In view of (9) and by applying Lemma 7.1(a) of Barlow and Proschan [1] we have

$$E(X_2^*) - E(X_1^*) = \int_0^\infty m(w) \, d\left[H_2(w) - H_1(w)\right] \geq 0.$$

The proof is now complete. ∎

**Theorem 6.3.**

Let $X$ be DLR (ILR). Then

$$\Theta_1 \leq_{RH} \Theta_2 \Rightarrow X_1^* \leq_{RH} (\geq_{RH}) X_2^*.$$

**Proof.**

We only prove the theorem for the case where $X$ is DLR. The other case is similar. Note that $X_1^* \leq_{RH} X_2^*$ if and only if, for all $0 \leq x_1 \leq x_2$, $F_1^*(x_2)F_2^*(x_1) \leq F_2^*(x_2)F_1^*(x_1)$. Because $\Theta_1$ and $\Theta_2$ are taken to be independent, we have for all $0 \leq x_1 \leq x_2$,

$$E\left\{\left(1 - [\bar{F}(x_2 + \Theta_2) / \bar{F}(\Theta_2)]\right) \times \left(1 - [\bar{F}(x_1 + \Theta_1) / \bar{F}(\Theta_1)]\right)\right\}$$

$$\geq E\left\{\left(1 - [\bar{F}(x_1 + \Theta_2) / \bar{F}(\Theta_2)]\right) \times \left(1 - [\bar{F}(x_2 + \Theta_1) / \bar{F}(\Theta_1)]\right)\right\}. \quad (10)$$

Let us define

$$\phi_1(\theta_1, \theta_2) = \left(1 - [\bar{F}(x_2 + \theta_1) / \bar{F}(\theta_1)]\right) \times \left(1 - [\bar{F}(x_1 + \theta_2) / \bar{F}(\theta_2)]\right),$$

and

$$\phi_2(\theta_1, \theta_2) = \left(1 - [\bar{F}(x_2 + \theta_2) / \bar{F}(\theta_2)]\right) \times \left(1 - [\bar{F}(x_1 + \theta_1) / \bar{F}(\theta_1)]\right),$$

where $0 \leq x_1 \leq x_2$ are fixed. Because $X$ is DLR thus Theorem 4.1 (i) applicable and in the model of (3) we can say that $X^*$ and $\Theta$ are PLRD. For all $0 \leq \theta_1 \leq \theta_2$, Shaked [15] proved that this means that $(X^* \mid \Theta = \theta_1) \leq_{LR} (X^* \mid \Theta = \theta_2)$, and because the likelihood ratio order implies the RHR order, we develop that $(X^* \mid \Theta = \theta_1) \leq_{RH} (X^* \mid \Theta = \theta_2)$, for all $0 \leq \theta_1 \leq \theta_2$, i.e., the function $k$ given by

$$k(\theta) = \left(1 - [\bar{F}(x_2 + \theta) / \bar{F}(\theta)]\right) / \left(1 - [\bar{F}(x_1 + \theta) / \bar{F}(\theta)]\right),$$

is increasing in $\theta > 0$, for all $0 \leq x_1 \leq x_2$. Hence, for all $0 \leq \theta_1 \leq \theta_2$,

$$\Delta\phi_{21}(\theta_1, \theta_2) = \phi_2(\theta_1, \theta_2) - \phi_1(\theta_1, \theta_2)$$

$$= \left(1 - [\bar{F}(x_2 + \theta_2) / \bar{F}(\theta_2)]\right) \times \left(1 - [\bar{F}(x_1 + \theta_1) / \bar{F}(\theta_1)]\right)$$

$$- \left(1 - [\bar{F}(x_2 + \theta_1) / \bar{F}(\theta_1)]\right) \times \left(1 - [\bar{F}(x_1 + \theta_2) / \bar{F}(\theta_2)]\right)$$

$$\geq 0.$$

Evidently, $\Delta\phi_{21}(\theta_1, \theta_2) = -\Delta\phi_{21}(\theta_2, \theta_1)$, for all $0 \leq \theta_1 \leq \theta_2$. It follows that

$$\Delta\phi_{21}(\theta_1, \theta_2) = \left(1 - [\bar{F}(x_1 + \theta_1) / \bar{F}(\theta_1)]\right) \times$$

$$\left\{\left(1 - [\bar{F}(x_2 + \theta_2) / \bar{F}(\theta_2)]\right) - k(\theta_1) \times \left(1 - [\bar{F}(x_1 + \theta_2) / \bar{F}(\theta_2)]\right)\right\},$$

is decreasing in $\theta_1$, for each $\theta_2$ such that $\theta_1 \leq \theta_2$. By applying the assumption that $\Theta_1 \leq_{RH} \Theta_2$ to the result of Theorem 1.B.48 of Shaked and Shanthikumar [5] we get the inequality given in (10) and hence the proof is completed. ∎

In the following counterexample we show that the conditions that $X$ is DLR (ILR) and that $X$ is DFR (IFR) cannot be dropped in the Theorems 6.1 - 6.3.

**Counterexample 6.1.**

Let $X$ have the sf $\bar{F}(t) = 1/(1 + t^2), t \geq 0$. Then, $X$ is not DFR (IFR) and because DLR (ILR) ia a subclass of DLR (ILR) hence $X$ is also not DLR (ILR). Suppose that $\Theta_1$ and $\Theta_2$ have the pdf's $h_1(\theta) = 4/\pi(1 + \theta^2)^2, \theta \geq 0$, and $h_2(\theta) = 2/\pi(1 + \theta^2), \theta \geq 0$, respectively. We observe that $\Theta_1 \leq_{LR} \Theta_2$ and thus $\Theta_1 \leq_{HR} (\leq_{RH})[\leq_{ST}]\Theta_2$. According to (3), the sf's of $X_1^*$ and $X_2^*$ are, respectively, obtained as

$$\bar{F}_1^*(x) = 4\,(\pi)^{-1} \left\{ [\pi - arctan(x)] \times [4 + x^2]^{-1} - [\ln(1 + x^2)] \times [4x(1+x^2)]^{-1} \right\}, \ x \geq 0,$$

and

$$\bar{F}_2^*(x) = 1 - 2\,(\pi)^{-1} arctan(x), \ x \geq 0.$$

It is seen that $\bar{F}_2^*(x) - \bar{F}_1^*(x)$ has a change of sign for $x \geq 0$. This means that $X_1^*$ and $X_2^*$ are not ordered in the usual stochastic ordering and hence they are not ordered in the HR, RHR and LR orders.

In the rest of this section, we consider the following model. Let $X_i$ have the sf $\bar{F}_i$ and let $X_i^*$ have the sf

$$\bar{F}_i^*(x) = E\left[\bar{F}_i(x + \Theta) \,/\, \bar{F}_i(\Theta)\right], \text{ for all } x \geq 0, \tag{11}$$

for $i = 1, 2$. Hence, we have two mixture models with a common random age variable $\Theta$ and different baseline variables $X_1$ and $X_2$. In the sequel, assume that $X_1$ and $X_2$ have pdf's $f_1$ and $f_2$, respectively.

**Theorem 6.4.**

Let $X_1$ be DLR and let $(\Theta \mid X_1^* = x) \leq_{LR} (\Theta \mid X_2^* = x)$, for all $x \geq 0$. Then

$$X_1 \leq_{LR} X_2 \ \Rightarrow \ X_1^* \leq_{LR} X_2^*.$$

**Proof.**

Let $f_i^*$ denote the pdf of $X_i^*$ associated with (11) and let $f_i(\cdot \mid \theta)$ denote the conditional pdf of $(X_i^* \mid \Theta = \theta)$, for $i = 1, 2$. We have

$$f_2^*(x) \,/\, f_1^*(x) = \left\{ E\left[f_2(x + \Theta) \,/\, \bar{F}_2(\Theta)\right]\right\} \,/\, \left\{ E\left[f_1(x + \Theta) \,/\, \bar{F}_1(\Theta)\right]\right\}$$

$$= \frac{\int_0^\infty \left\{ [f_2(x + w) \,/\, f_1(x + w)] \left[\bar{F}_1(w) \,/\, \bar{F}_2(w)\right] \left[f_1(x + w) \,/\, \bar{F}_1(w)\right]\right\} \, dH(w)}{\int_0^\infty \left[f_1(x + w) \,/\, \bar{F}_1(w)\right] \, dH(w)}$$

$$= E[\phi(x, W)],$$

where

$$\phi(x, w) = [f_2(x + w) \,/\, f_1(x + w)] \times \left[\bar{F}_1(w) \,/\, \bar{F}_2(w)\right],$$

$W$ is a nonnegative random variable with the following pdf

$$h(w \mid x) = \left[\int_0^\infty [f_1(x + w) \,/\, \bar{F}_1(w)] \, dH(w)\right]^{-1} \times \left\{[f_1(x + w)h(w)] \,/\, \bar{F}_1(w)\right\}, \ w, x \geq 0.$$

Denote by $\pi_i(\cdot \mid x)$ the conditional pdf of $(\Theta \mid X_i^* = x)$, for each $i = 1, 2$. Because $(\Theta \mid X_1^* = x) \leq_{LR} (\Theta \mid X_2^* = x)$, for all $x \geq 0$, thus

$$[\pi_2(\theta \mid x) \,/\, \pi_1(\theta \mid x)] \ = \ [f_2(x \mid \theta) f_2^*(x)] \,/\, [f_1(x \mid \theta) f_1^*(x)]$$

$$= \ \{f_2(x + \theta) \,/\, f_1(x + \theta)\} \times \{\bar{F}_1(\theta) \,/\, \bar{F}_2(\theta)\} \times \{f_2^*(x) \,/\, f_1^*(x)\}$$

is increasing in $\theta$, for any $x \geq 0$. Therefore, $\phi(x,w)$ is increasing in $w$, for any $x \geq 0$. In parallel, because $X_1 \leq_{LR} X_2$, thus $\phi(x,w)$ is increasing in $x$, for all $w \geq 0$. In a similar manner to the proof of Theorem 5.1, the assumption that $X_1$ is DLR yields $W$ is stochastically increasing in $x$. Hence Lemma 2.2 (i) of Misra and Van Der Meulen [13] gives $E[\phi(x,W)]$ is increasing in $x$, or equivalently $X_1^* \leq_{LR} X_2^*$. ∎

**Theorem 6.5.**

Let at least one of $X_1$ and $X_2$ be DFR and let $(\Theta \mid X_1^* > x) \leq_{ST} (\Theta \mid X_2^* > x)$, for all $x \geq 0$. Then
$$X_1 \leq_{HR} X_2 \Rightarrow X_1^* \leq_{HR} X_2^*.$$

**Proof.**

Suppose that $\Pi_i(\cdot \mid x)$ is used to denote the df of $(\Theta \mid X_i^* > x)$, for $i = 1, 2$. In view of (6) and because $X_1 \leq_{HR} X_2$, we can write, for all $x \geq 0$, that

$$r_{X_1^*}(x) - r_{X_2^*}(x) = E\left[r_{X_1}(x+\Theta) \mid X_1^* > x\right] - E\left[r_{X_2}(x+\Theta) \mid X_2^* > x\right]$$

$$= \int_0^\infty r_{X_1}(x+w)\, d\Gamma_1(w \mid x) - \int_0^\infty r_{X_2}(x+w)\, d\Gamma_2(w \mid x)$$

$$\geq \int_0^\infty r_{X_i}(x+w)\, d\left[\Gamma_1(w \mid x) - \Gamma_2(w \mid x)\right], \text{ for each } i = 1, 2. \quad (12)$$

We know by assumption that for at least one of $i = 1$ and $i = 2$, the random variable $X_i$ is DFR, i.e., $r_{X_i}(x+w)$ is decreasing in $w$, for all $x \geq 0$. On the other hand, $(\Theta \mid X_1^* > x) \leq_{ST} (\Theta \mid X_2^* > x)$, for all $x \geq 0$, implies that

$$\int_0^\theta d\left[\Pi_1(w \mid x) - \Pi_2(w \mid x)\right] \geq 0, \text{ for all } \theta \geq 0.$$

Finally, by applying Lemma 7.1(b) of Barlow and Proschan [1] to (12) we conclude the result. ∎

**Theorem 6.6.**

Let at least one of $X_1$ and $X_2$ be IMRL and let $(\Theta \mid X_1^* > x) \leq_{ST} (\Theta \mid X_2^* > x)$, for all $x \geq 0$. Then
$$X_1 \leq_{MRL} X_2 \Rightarrow X_1^* \leq_{MRL} X_2^*.$$

**Proof.**

Consider the notations introduced in Theorem 6.4. In view of (7) and because $X_1 \leq_{MRL} X_2$, we can write, for all $x \geq 0$, that

$$m_{X_2^*}(x) - m_{X_1^*}(x) = E\left[m_{X_2}(x+\Theta) \mid X_2^* > x\right] - E\left[m_{X_1}(x+\Theta) \mid X_1^* > x\right]$$

$$= \int_0^\infty m_{X_2}(x+w)\, d\Pi_2(w \mid x) - \int_0^\infty m_{X_1}(x+w)\, d\Pi_1(w \mid x)$$

$$\geq \int_0^\infty m_{X_i}(x+w)\, d\left[\Pi_2(w \mid x) - \Pi_1(w \mid x)\right], \text{ for } i = 1, 2. \quad (13)$$

We know by assumption that for at least one of $i = 1$ and $i = 2$, the random variable $X_i$ is IMRL, i.e., $m_{X_i}(x+w)$ is increasing in $w$, for all $x \geq 0$. On the other hand, $(\Theta \mid X_1^* > x) \leq_{ST} (\Theta \mid X_2^* > x)$, for all $x \geq 0$, implies that

$$\int_\theta^\infty d\left[\Gamma_2(w \mid x) - \Gamma_1(w \mid x)\right] \geq 0, \text{ for all } \theta \geq 0.$$

On applying Lemma 7.1(a) of Barlow and Proschan [1] to (13) the result follows. ∎

## 7  Summary and concluding remarks

Based on the concept of the mixture distribution, a new extended mixture model of the family of residual lifetime distributions $\{\bar{F}(x \mid \theta) = \bar{F}(x + \theta) / \bar{F}(\theta) \mid \theta > 0\}$ with respect to the mixing distribution $H$ of $\theta$ was introduced and studied. The average residual life variable was denoted by $X^*$ and the random age variable was denoted by $\Theta$. Several bivariate dependence properties such as PLRD (NLRD), RCSI (RCSD), and SI (SD) between $X^*$ and $\Theta$ were characterized via some well-known aging classes. We established some characterizations of various aging properties by making several stochastic orders between $X$ and $X^*$. In addition, we provide various closure properties of the new model with respect to some stochastic orders like upshifted likelihood ratio, upshifted hazard rate, and upshifted mean residual life orders and with respect to some aging classes such as DLR, DFR, and IMRL. We investigated that how stochastic orders between two random age variables are translated to stochastic orders of the associated average residual life variables. Finally, we provide some conditions under which stochastic orders between two variables are translated to stochastic orders between their average residual life variables. Our results provide new concepts and applications in reliability, statistics and operations research. Further properties and applications of the new model can be considered in the future of this research.

**ACKNOWLEDGEMENTS**

The authors would like to thank the Editor and anonymous reviewers for their valuable comments and suggestions, which were helpful in improving the paper. The authors also would like to extend their sincere appreciation to the Deanship of Scientific Research at King Saud University for its funding this Research Group NO (RG-1435-036).